\documentclass[pra,preprint,showpacs,superscriptaddress]{revtex4}

\usepackage{amsmath}
\usepackage{epsfig}

\begin{document}

\title{Packet narrowing and quantum entanglement in photoionization
and photodissociation}
\author{M. V. Fedorov}
\email{fedorov@ran.gpi.ru}
\author{M. A. Efremov}
\author{A. E. Kazakov}
\affiliation{General Physics Institute, Russian Academy of
Sciences, 38 Vavilov st, Moscow, 119991 Russia}
\author{K. W. Chan}
\email{kwchan1@pas.rochester.edu}
\affiliation{{Center for Quantum Information and Department of Physics
and Astronomy,} \\ {University of Rochester, Rochester, NY 14627 USA}}
\author{C. K. Law}
\affiliation{Department of Physics, The Chinese University of Hong
Kong, NT, Hong Kong SAR, China}
\author{J. H. Eberly}
\affiliation{{Center for Quantum Information and Department of Physics
and Astronomy,} \\ {University of Rochester, Rochester, NY 14627 USA}}

\date{\today}

\begin{abstract}
The narrowing of electron and ion wave packets in the process of
photoionization is investigated, with the electron-ion recoil fully
taken into account. Packet localization of this type is directly  related
to entanglement in the joint quantum state of electron and ion, and to
Einstein-Podolsky-Rosen localization. Experimental observation of such
packet-narrowing effects is suggested via coincidence registration by two
detectors, with a fixed position of one and varying position of the
other.  A similar effect, typically with an enhanced degree of
entanglement, is shown to occur in the  case of photodissociation of
molecules.
\end{abstract}


\maketitle

\section{Introduction}

In the course of photoionization, a photoelectron is ejected and the ion
recoils, being constrained both by the conservation of momentum and
energy and by the condition of the original atom. In principle this
initial condition includes all aspects of the atom's internal and
center of mass states, but here we will focus for greatest clarity on
an atom in its ground electronic state with a center of mass wave
packet determined by whatever localizes the atom in the region of the
photoionization.  Although photoionization has been treated
repeatedly in weak and strong fields (see, for example,
Refs.~\cite{Gottfried} and~\cite{MVF}), the focus has been mainly on
cross sections and the dynamics of the particles, and there has been
little discussion of the nature of the joint quantum state of the
breakup fragments. This quantum state is entangled and it is closely
related to questions of fundamental interest because
the fragmentation process is exactly the one used by Einstein, Podolsky
and Rosen (EPR) \cite{EPR} to illustrate Einstein's position regarding
limitations of quantum theory. We identify the amount of state
entanglement by examining the relative localization of the wavepackets of
the electron and ion, under the simplifying assumptions that the
incident photon
momentum and the post-breakup Coulomb interaction can be neglected. The
different time regimes for entanglement are identified.

Due to the finite mass of the particles, the wave function of the
system changes drastically between the initial stage and the
long-time limit. We find suitable measures of the entanglement of
the electron and ion that are connected with their packet widths
in position space, specifically the coincidence width and the
single-particle width. In view of other treatments of breakup
\cite{JHE2}, these localization measures give an alternative view
of entanglement and reveal new channels for achieving high degrees
of entanglement. Our choice of measure also identifies
entanglement ``control parameters" for comparison with those that
have been advanced in previous studies of both photon-atom
\cite{JHE} and photon-photon \cite{PDC} wave functions, and
through conditional localization it is related to formal photonic
analogues \cite{Reid} of the EPR discussion and bimolecular
breakup as a route to matter wave entanglement \cite{OK}. More
importantly, these widths are experimentally measurable entities.
Finally, we discuss similar issues for dissociation of a diatomic
molecule and explain the most significant differences in the
results.

\section{Photoionization}

Let an atom, originally in its ground state, be photoionized by a
light field
\begin{equation}
\label{field}
     \vec{\mathcal{E}} (t)=\vec{\mathcal{E}}_0\,\sin(\omega t),
\end{equation}
where $\hbar\omega>|E_0|$ and $E_0$ is the ground state energy.
It should be noted that by using the dipole approximation and
ignoring the term $\vec{k}\cdot\vec{r}$ in the argument in
Eq.~({\ref{field}}) we are ignoring all recoil effects due
to absorption of the photon momentum $\hbar\vec{k}$.
This approximation is quite reasonable because there is another much
stronger mechanism giving rise to recoil. In the process of
photoionization an atomic electron acquires an energy
$\sim\hbar\omega$ and hence a momentum $\sim\sqrt{m\hbar\omega}$,
and the ion gets the same momentum (with the opposite sign),
and this momentum is much larger than $\hbar k=\hbar\omega/c$.
This is in contrast to the problems of entanglement in spontaneous
photon emission of excited atoms and Raman scattering~\cite{JHE}.

To describe such a process with atomic recoil and with an
initial wave-packet distribution of the atomic center of mass, let
us begin from the Schr\"{o}dinger equation for two
particles - electron and ion - in the field.
Traditionally, to separate variables in such an
equation, we use the relative (rel) and center-of-mass (cm)
position and momentum vectors~\cite{LL}
\begin {equation}
\label{rel-cm}
     \begin{array}{ll}
         \vec{r}_\text{rel}=\vec{r}_e-\vec{r}_i,
         \quad
         &\vec{r}_\text{cm}=\displaystyle\frac{m_e \vec{r}_e+
m_i\vec{r}_i}{M},
         \\
         \vec{p}_\text{rel}=
\displaystyle\frac{m_i\vec{p}_e-m_e\vec{p}_i}{M},
         \quad
         &\vec{p}_\text{cm}= \vec{p}_e+\vec{p}_i,    \end{array}
\end{equation}
where $\vec{r}_e$ and $\vec{r}_i$ are the electron and ion
position vectors, $\vec{p}_e=-i\hbar\partial/\partial\vec{r}_e$
and $\vec{p}_i=-i\hbar\partial/\partial\vec{r}_i$ are their
momenta, and $m_e$ and $m_i$ are their masses, with $M=m_e+m_i$. It
is worth noting that the ``mixed'' coordinate-momentum variable pairs
$\vec{r}_\text{rel}$ and $\vec{p}_\text{cm}$, as well as
$\vec{r}_\text{cm}$ and $\vec{p}_\text{rel}$, each have zero
commutator. For example, $[\vec{r}_\text{rel}, \vec{p}_\text{cm}] =
0$. For this reason one can call them EPR pairs, recalling the famous
discussion of Einstein, Podolsky and Rosen~\cite{EPR}.

The Schr\"{o}dinger equation takes the form
\begin{equation}
\label{Schr}
     i\hbar\frac{\partial\Psi}{\partial t}
     = \left\{\frac{\vec{p}_\text{cm}^{\;2}}{2M}
     + \frac{\vec{p}_\text{rel}^{\;2}}{2\mu} - \frac{e^2}{r_\text{rel}}
     + e\vec{r}_\text{rel}\cdot\vec{\mathcal{E}}_0\sin(\omega
t)\right\} \Psi,
\end{equation}
where $\mu=m_e m_i/M$ is the reduced mass.
Because we have made the dipole approximation, the
variables $\vec{r}_\text{rel}$ and $\vec{r}_\text{cm}$ in Eq.
(\ref{Schr}) are separated, and its solution is a product of
functions depending on these two variables separately:
\begin{equation}
\label{product}
     \Psi(\vec{r}_\text{rel},\vec{r}_\text{cm},t)
     = \Psi_\text{cm}(\vec{r}_\text{cm},t)
\times\Psi_\text{rel}(\vec{r}_\text{rel},t) ,
\end{equation}
where the equations of motion of $\Psi_\text{cm}(\vec{r}_\text{cm},t)$ and
$\Psi_\text{rel}(\vec{r}_\text{rel},t)$ are
\begin{equation}
\label{Schr-cm}
     i\hbar\frac{\partial\Psi_\text{cm}}{\partial t}
     = \frac{\vec{p}_\text{cm}^{\;2}}{2M}\,\Psi_\text{cm}
\end{equation}
and
\begin{equation}
\label{Schr-rel}
     i\hbar\frac{\partial\Psi_\text{rel}}{\partial t}
     = \left\{\frac{\vec{p}_\text{rel}^{\;2}}{2\mu}
     - \frac{e^2}{r_\text{rel}} + e\vec{r}_\text{rel}
     \cdot\vec{\mathcal{E}}_0\sin(\omega t)\right\} \Psi_\text{rel} .
\end{equation}
We note that the factorization shown in~(\ref{product}) is far from the same as
factorization in the particle variables $\vec{r}_e$ and $\vec{r}_i$.  That is,
the electron and ion are quantum entangled in the state given
in~(\ref{product}).

Let us assume that the initial atomic center-of-mass wave function
is given by a Gaussian wave packet with width $\Delta r_\text{cm}^{(0)}$:
\begin{equation}
\label{wp-init}
     \Psi_\text{cm}(\vec{r}_\text{cm},t=0)
     = \frac{1}{(2\pi)^{\frac{3}{4}}\left[\Delta
r_\text{cm}^{(0)}\right]^{\frac{3}{2}}}
     \exp\left(-\frac{\vec{r}_\text{cm}^{\;2}}{4\left[\Delta
r_\text{cm}^{(0)}\right]^2}\right) .
\end{equation}
Then, as is well known \cite{MVF}, the time-dependent solution
of Eq.~(\ref{Schr-cm}) has the form of a spreading wave packet
such that
\begin{equation}
\label{wp-spr}
     |\Psi_\text{cm}(\vec{r}_\text{cm},t)|^2
     = \frac{1}{(2\pi)^{\frac{3}{2}}\left[\Delta r_\text{cm}(t)\right]^3}
     \exp\left(-\frac{\vec{r}_\text{cm}^{\;2}}{2\left[\Delta
r_\text{cm}(t)\right]^2}\right),
\end{equation}
where $\Delta r_\text{cm}(t)$ is the time-dependent width of the
center-of-mass wave packet (\ref{wp-spr})
\begin{equation}
\label{spr-width}
     \Delta r_\text{cm}(t)
     = \left\{\left[\Delta r_\text{cm}^{(0)}\right]^2
     + \frac{\hbar^2\,t^2}{4M^2 \, \left[\Delta
     r_\text{cm}^{(0)}\right]^2}\right\}^{1/2}
     \approx\left\{
     \begin{array}{cr}
     \Delta r_\text{cm}^{(0)},&\quad t\ll t_{{\rm spr}}^{({\rm
     cm})}\\
     \,&\,\\
     \displaystyle\frac{\hbar\,t}{2M \, \Delta
r_\text{cm}^{(0)}},&\quad t\gg t_{{\rm spr}}^{({\rm
     cm})}
     \end{array}
     \right.
\end{equation}
and $t_\text{spr}^\text{(cm)}$ is its spreading time, $t_{{\rm
spr}}^{({\rm cm})}=2M\left[\Delta r_{{\rm
cm}}^{(0)}\right]^2/\hbar$. At $t\gg t_{{\rm spr}}^{({\rm cm})}$
the width $\Delta r_\text{cm}(t)$ grows linearly and the velocity
of spreading equals to $v_{{\rm spr}}^{({\rm cm})}=\hbar/2M\Delta
r_{{\rm cm}}^{(0)}$.

  Under the conditions of interest here, the solution of
Eq.~(\ref{Schr-rel}) is only a little bit more complicated.  The
initial wave function of the relative motion is taken to be the
hydrogen ground-state $1s$ wave function
\begin{equation}
\label{H-1s}
     \Psi_\text{rel}(\vec{r}_\text{rel},t=0)
     \ = \ \psi_{1s}
     \ \equiv \ R_{10}(r_\text{rel}) \, Y_{00},
\end{equation}
where $R_{10}(r_\text{rel})$ is the hydrogen radial wave function for
the principal quantum number $n=1$ and angular momentum $l=0$, and
$Y_{00}=1/\sqrt{4\pi}$ is the spherical function for $l=m_l=0$.

We assume a sufficiently high photon energy $\hbar\omega$ to ignore bound-bound
transitions.  Then the time-dependent wave function obeying
Eq.~(\ref{Schr-rel})
can be presented in the form
\begin{equation}
\label{superposition}
     \Psi_\text{rel}(\vec{r}_\text{rel},t)
     = C_0(t) \, \psi_{1s} + e^{-i\omega t} \int_0^{\infty}dE \
C_E(t) \,\psi_{Ep},
\end{equation}
where $\psi_{Ep}$ is the field-free wave function of the continuous spectrum
with $l=1,\,m_l=0$:
\begin{equation}
\label{H-Ep}
     \psi_{Ep}(\vec{r}_\text{rel})
     = R_{E1}(r_\text{rel}) \, Y_{10}(\cos\theta_\text{rel}),
\end{equation}
in which $R_{E1}(r_\text{rel})$ is the radial wave function for energy $E$
and angular momentum $l=1$, $Y_{10}=\sqrt{3/4\pi}\,\cos\theta_\text{rel}$,
and $\theta_\text{rel}$ is the angle between the vectors $\vec{\mathcal{E}}_0$
and $\vec{r}_\text{rel}$.

With multi-photon processes ignored, the equations of motion for the
probability amplitudes $C_0(t)$ and $C_E(t)$ in the rotating-wave
approximation, following directly from Eq.~(\ref{Schr-rel}), are given by
\begin{subequations}
     \begin{eqnarray}
         \hspace{-10mm} &&
         i\hbar\dot{C}_0(t)-E_0\,C_0(t)
         = - \int_0^\infty dE\,
\frac{\vec{d}_{0E}\cdot\vec{\mathcal{E}_0}}{2}\,C_E(t),
\label{C0-CE-eq-a}
         \\ \hspace{-10mm} &&
         i\hbar\dot{C}_E(t)-(E-\hbar\omega)\,C_E(t)
         = - \frac{\vec{d}_{E0}\cdot\vec{\mathcal{E}}_0}{2}\,C_0(t),
\label{C0-CE-eq-b}
     \end{eqnarray}
\end{subequations}
where $\vec{d}_{E0}=(\vec{d}_{0E})^*$ are the bound-free dipole matrix
elements of the atom, and we consider the case of a pulse with rectangular
envelope, which means that the interaction is turned on suddenly at $t=0$.

With the help of adiabatic elimination of the continuum~\cite{MVF},
Eq.~(\ref{C0-CE-eq-a}) can be reduced to a much simpler form:
\begin{equation}
\label{C0-rate-eq}
     i\hbar\dot{C}_0(t)-\left(E_0- i\hbar\gamma_I\right)\,C_0(t) = 0 ,
\end{equation}
in which the amplitude decay rate $\gamma_I$ is half the Fermi-Golden Rule
rate of ionization:
\begin{equation}
\label{FGR}
     2 \gamma_I \equiv \frac{dw_I}{dt}
     = \frac{2\pi}{\hbar}\left.\left|\left\langle E\left|
     \frac{\vec{d}\cdot\vec{\mathcal{E}}_0}{2}\right|0\right
     \rangle\right|^2\right|_{E=E_0+\omega} .
\end{equation}
The solution satisfying the initial condition $C_0(0)=1$ is
\begin{equation}
\label{C0-solution}
     C_0(t) = \exp\left(-\frac{i}{\hbar}E_0t-\gamma_I t\right) .
\end{equation}
With this function substituted into the right-hand side of
Eq.~(\ref{C0-CE-eq-b}), the equation for $C_E(t)$ can be easily solved
to give, with the initial condition $C_E(0)=0$:
\begin{eqnarray}
\label{CE-solution}
     && \hspace{-10mm}
     C_E(t)
     =
\frac{1}{2}\frac{\vec{d}_{E0}\cdot\vec{\mathcal{E}}_0}{E-E_0-\hbar\omega+i\gamma_I}
     \times \nonumber \\
     &&  \hspace{-10mm}
     \Bigg\{\exp\left[-\left(\frac{i E_0}{\hbar} +\gamma_I\right)t\right]-
     \exp\left[-i\left(\frac{E}{\hbar}-\omega\right)t\right]\Bigg\}.
\end{eqnarray}
At times $t \gg \gamma_I^{-1}$ both $C_0(t)$ in Eq.~(\ref{C0-solution}) and
the first exponential term in Eq.~(\ref{CE-solution}) vanish.
As a result the wave function $\Psi_\text{rel}$ describing relative
motion takes the form
\begin{eqnarray}
\label{Psi-rel-solu}
     \Psi_\text{rel}(\vec{r}_\text{rel}, t)
     &=& \frac{-\sqrt{3}}{4\sqrt{\pi}} \cos\theta_\text{rel}
     \int_0^\infty dE\,\,R_{E1}(r_\text{rel}) \nonumber \\
     & & \hspace{-8mm}
     \times\exp\left(-\frac{i}{\hbar}\,E\,t\right)\,
\frac{\vec{d}_{E0}\cdot\vec{\mathcal{E}}_0}{E-E_0-\hbar\omega+i\hbar\gamma_I}
.
\hspace{8mm}
\end{eqnarray}
We assume that the laser frequency $\omega$ and hence the
energy $E\sim E_0+\hbar\omega$ are high enough so that
the radial function $R_{E1}(r_\text{rel})$ is approximated
by the well known high-energy field-free expression for the
Coulomb radial wave function~\cite{LL}:
\begin{equation}
\label{RE-high-E}
     R_{E1}(r)
     \approx \sqrt{\frac{2\mu}{\pi k}}\;\frac{1}{\hbar r}
     \cos\left(kr+\frac{1}{k a_0}\ln(2kr)+\delta_1\right),
\end{equation}
where $k=\sqrt{2\mu E}/\hbar$, $\delta_1$ is the Coulomb scattering phase
for $l=1$, and $a_0=\hbar^2/\mu e^2$ is the Bohr radius.

When the photoelectrons have energy far above the continuum threshold,
we have $\hbar\gamma_I\ll E\sim E_0+\omega$.  In this way the lower limit
of the integration over $E$ in Eq.~(\ref{Psi-rel-solu}) can be
replaced by $-\infty$.
The energy $E$ is approximated by $E_*\equiv E_0+\hbar\omega$ in all the
pre-exponential factors except the denominator on the right-hand side of
Eq.~(\ref{Psi-rel-solu}).  Also both the scattering phase $\delta_1$ and
logarithmic term in the argument of cosine in Eq.~(\ref{RE-high-E})
are neglected,
and the factor $k$ in the product $kr$ is expanded in powers of $E-E_*$,
\textit{viz}, $k\approx k_*+(E-E_*)/\hbar v$, where $k_*=\sqrt{2\mu E_*}/\hbar$
and $v=\sqrt{2E_*/\mu}=\hbar k_*/\mu$ is the velocity of the relative motion.
Then the integral over $E$ can then be evaluated by the residue method, giving
\begin{eqnarray}
\label{Psi-residue-int}
     \hspace{-1mm}
     \Psi_\text{rel}
     &=& \frac{i\sqrt{6}}{4\sqrt{\hbar v}}
     \left(\vec{d}_{E_*0}\cdot\vec{\mathcal{E}}_0\right)
     \exp\left(-i\frac{E_0 t}{\hbar} + i k_*r_\text{rel}\right)
     \nonumber \\
     \hspace{-1mm}
     & & \hspace{-3mm}
     \times \frac{\cos\theta_\text{rel}}{r_\text{rel}}
     \exp\left[-\gamma_I\left(t-\frac{r_\text{rel}}{v}\right)\right]
     \theta (vt-r_\text{rel}).
     \quad
\end{eqnarray}
This equation describes a spherical wave packet in $r_\text{rel}$
with an angular modulation determined by the factor
$\cos\theta_\text{rel}$, propagating in the direction of growing
$r_\text{rel}$ with velocity $v$, having a sharp edge at
$r_\text{rel}=vt$ and an exponentially falling tail at
$r_\text{rel}<vt$. The radial width of the wave packet
(\ref{Psi-residue-int}) is $v/2\gamma_I$.

\section{Evolution of the relative-motion wave function}

\begin{figure}[!htb]
     \centering
     \centering\epsfig{file=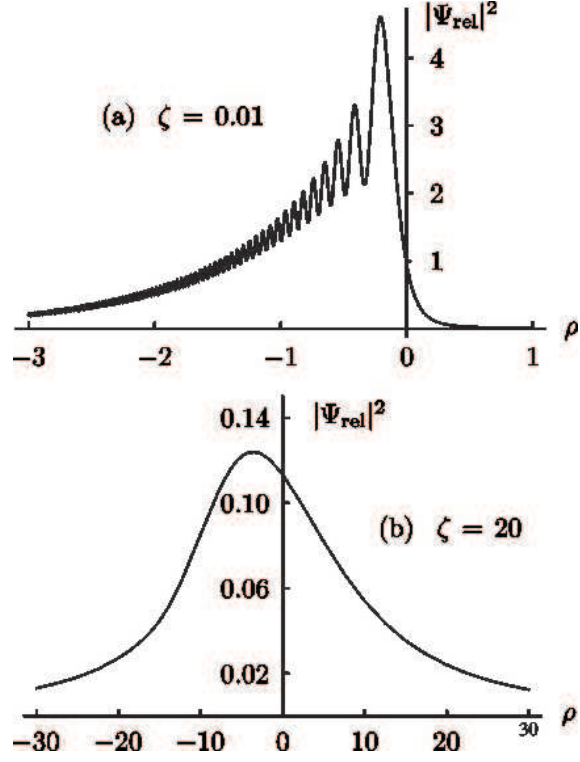, width=7.5cm}
     \caption{The relative-motion probability density $|\Psi_\text{rel}|^2$
     (\ref{Psi-rel-spreading}) in dependence on $\rho=
     (r_\text{rel}-vt)/\Delta r_\text{rel}^{(0)}$ at (a) $\zeta=0.01$
     and (b) $\zeta=20$.}
\label{fig1}
\end{figure}

Although we have assumed that the time $t$ exceeds the total ionization
time ($t>\gamma_I^{-1}$), Eq.~(\ref{Psi-residue-int}) still describes the
initial stage for the relative-motion wave packet evolution after ionization.
In this sense $v/2\gamma_I$ is the initial width of the relative-motion wave
packet $|\Psi_\text{rel}|^2$, which we denote with
$\Delta r_\text{rel}^{(0)} \equiv v/2\gamma_I$. This width
can change later due to dispersion.  To describe such a spreading
effect, we can extend our series expansion of the function $k(E)$ up
to second order in $E-E_*$: $k\approx k_* +(E-E_*)/\hbar
v-(E-E_*)^2/2\hbar \mu v^3$. This gives rise to an additional
factor in the integral over the energy $E$:
$\exp\left\{-i\,E^2r_\text{rel}/2\hbar\mu v^3\right\}$.
To keep the possibility of integration by the residue method, we have
to use the Fourier transformation of this factor
\begin{eqnarray*}
     && \exp\left\{-i\frac{E^2r_\text{rel}}{2\hbar\mu v^3}\right\} \\
     && = \sqrt{\frac{\mu v^3}{2\pi i\hbar r_\text{rel}}}
     \int_{-\infty}^\infty d\tau \ \exp\left\{\frac{i}{\hbar}
     \left[E\tau+\frac{\mu v^3\tau^2}{2r_\text{rel}}\right]\right\}.
\end{eqnarray*}
With this representation we first carry out the integration
over $E$ (by the residue method), and then the one over $\tau$.
The result is
\begin{equation}
     \label{Psi-rel-spreading}
     \begin{split}
         \hspace{-2mm}
         |\Psi_\text{rel}(r_\text{rel},t)|^2
         = \frac{3}{16\pi \Delta r_\text{rel}^{(0)}}
         \frac{\cos^2\theta_\text{rel}}{r_\text{rel}^2}
         \exp\left(\frac{r_\text{rel}-vt}{\Delta
r_\text{rel}^{(0)}}\right)
         \\[2mm]
         \hspace{-2mm}
         \times\left|1-{\rm
Erf}\left[\sqrt{\frac{i}{2}}\left(\frac{\sqrt{\zeta}}{2}
         -
\frac{i}{\sqrt{\zeta}}\,\frac{r_\text{rel}-vt}{\Delta
r_\text{rel}^{(0)}}
         \right)\right]\right|^2,
     \end{split}
\end{equation}
where ${\rm Erf}$ is the error function, and we have defined
\begin{equation}
     \zeta
     \equiv \frac{\hbar r_\text{rel}}{v\mu (\Delta r_\text{rel}^{(0)})^2}
     \equiv \frac{r_\text{rel}}{v t_\text{spr}} ,
\end{equation}
so that $t_\text{spr}^{({\rm rel})}=\mu (\Delta
r_\text{rel}^{(0)})^2/\hbar$ is the spreading time of the
relative-motion wave packet. As the value of $|\Psi_\text{rel}|^2$
is concentrated around $r_\text{rel} \approx vt$, by putting
$r_\text{rel}\approx vt$ in the definition of the
parameter~$\zeta$, we get $\zeta=t/t_\text{spr}^{({\rm rel})}$. In
this form the meaning of $\zeta$ is obvious: it is the time after
ionization measured in units of the spreading time of the
relative-motion wave packet. In Figs.~\ref{fig1}(a)
and~\ref{fig1}(b) the function $|\Psi_\text{rel}|^2$ is plotted in
its dependence on $\rho\equiv (r_\text{rel}-vt)/\Delta
r_\text{rel}^{(0)}$ at small and large $\zeta$'s respectively.

In the small-spreading regime ($\zeta\ll 1$) $|\Psi_\text{rel}|^2$
returns to the form of Eq.~(\ref{Psi-residue-int}), but with
additional oscillations on the left wing and a slightly smoothed
right wing as compared to the step function jump of
Eq.~(\ref{Psi-residue-int}). In the large-spreading regime ($\zeta\gg
1$) $|\Psi_\text{rel}|^2$ takes a Lorentzian shape:
\begin{equation}
\label{Lorentz}
     |\Psi_\text{rel}|^2
     = \frac{3}{8\pi^2} \frac{\cos^2\theta_\text{rel}}{r_\text{rel}^2}
     \frac{\Delta r_\text{rel}(t)}{(r_\text{rel}-vt)^2 +
     \frac{1}{4}\left[\Delta r_\text{rel}(t)\right]^2},
\end{equation}
where
\begin{equation}
\label{Detlta-rel-spr}
     \Delta r_\text{rel}(t)=\zeta\Delta r_\text{rel}^{(0)}
     = \frac{t}{t_\text{spr}}\Delta r_\text{rel}^{(0)}
     = v_\text{spr}t
\end{equation}
and $v_\text{spr}\equiv\Delta r_\text{rel}^{(0)}/t_\text{spr}=
\hbar/\mu\Delta r_\text{rel}^{(0)}$. Altogether, at small
and large~$\zeta$, the time-dependent width of the relative-motion
wave packet is given by
\begin{equation}
\label{Detlta-rel-tot}
     \Delta r_\text{rel}(t)
     = \left\{
     \begin{array}{ll}
         \Delta r_\text{rel}^{(0)}
         = \displaystyle\frac{v}{2\gamma_I}, \quad & t\ll t_{{\rm
spr}}^{({\rm rel})} \;(\zeta\ll 1); \\[5mm]
         v_\text{spr} t
         =\displaystyle\frac{\hbar t}{\mu \Delta r_\text{rel}^{(0)}}
         =\displaystyle\frac{2\hbar\gamma_I}{\mu v} t , \quad
& t\gg t_{{\rm spr}}^{({\rm rel})} \;(\zeta\gg 1).
         \end{array}
     \right.
\end{equation}
In spite of a difference between the Gaussian center-of-mass
(\ref{wp-spr} ) and relative-motion
[(\ref{Psi-residue-int}),(\ref{Psi-rel-spreading}),
(\ref{Lorentz})] wave packets, their widths behave similarly: in
dependence on $t$ they start from the initial values $\Delta
r_\text{cm}^{(0)}$ and $\Delta r_\text{rel}^{(0)}$, and at time
$t$ longer than the corresponding spreading time both $\Delta
r_\text{rel}(t)$ (\ref{Detlta-rel-tot}) and $\Delta
r_\text{cm}(t)$  (\ref{spr-width}) grow linearly. In both cases
the spreading time is inversely proportional to the squared
initial size and the velocity of spreading is inversely
proportional to the initial size to the first power. The only
qualitative difference concerns the mass of an objet: the total
mass $M$ of the center-of-mass wave function is substituted by the
reduced mass $\mu$ in the case of the relative-motion wave packet.

The relation between the center-of-mass and relative-motion wave
packet widths can change with time due to different spreading
velocities of these wave packets.  This makes the time evolution
of the electron-ion wave function rather complicated, and this
problem will be discussed separately in
Section~\ref{section:evolution}.

\section{Localization of the electron-ion wave packet and entanglement}

Before going further into details of time evolution, let us
discuss the entanglement effect. As we will show, in cases of
initially localized pairs of particles as in photoionization, and
where significant further interaction is absent, entanglement can
be evaluated by carrying out a series of localization
measurements. This has a close analog in earlier studies of
spontaneous photon emission with atom recoil \cite{JHE, JHE2,
SPIE} as well as in the measurement-induced localization and
entanglement discussed recently in a very different context by
Rau, et al. \cite{RDK}.

We will proceed by determining the dependence of entanglement on the
ratio of widths
\begin{equation}
\label{eq:eta}
     \eta(t) \equiv \frac{\Delta r_\text{cm}(t)}{\Delta r_\text{rel}(t)}
\end{equation}
where the time $t$ is taken as a parameter.
Note that in our treatment $\eta(t)$ is constrained
only by momentum and energy conservation (e.g., we ignore final-state
electron-ion Coulomb effects).  Here $\Delta r_\text{cm}(t)$
is of kinematic origin whereas $\Delta r_\text{rel}(t)$ is due to the dynamics
of the ionization process. We will see that $\eta(t)$ acts as the sole
control parameter for entanglement of the two-particle system.

In accord with Eq.~(\ref{product}), the product of the wave
functions Eqs.~(\ref{wp-spr}) and~(\ref{Psi-rel-spreading}) determines
the total wave function of the ion-electron system.  It should now be
considered as a function of ion and electron position vectors
\begin{equation}
\label{product-2}
     \Psi(\vec{r}_e,\vec{r}_i,t)
     = \Psi_\text{cm}\left(\frac{m_e\vec{r}_e+m_i\vec{r}_e}{M},t\right)
     \times\Psi_\text{rel}(\vec{r}_e-\vec{r}_i,t) ,
\end{equation}
showing that both $\Psi$ and its squared absolute
value are not factorable in the individual particle coordinates $\vec{r}_e$ and
$\vec{r}_i$. Such non-factorization defines quantum particle entanglement of
electron and ion.

\begin{figure}[!htb]
     \centering\epsfig{file=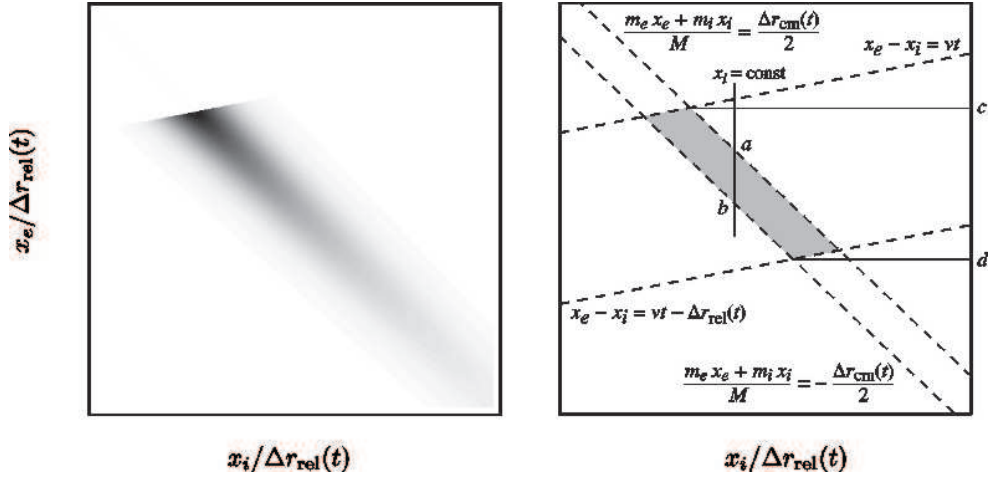, width=13cm}
     \caption{Views of the one-dimensional equivalent of
     $|\Psi|^2$ in Eq.~(\ref{product-2}).  Here we illustrate
     the relation between entanglement and distribution
     of particle positions.  High correlation between $x_e$ and $x_i$ can
     be attained only under the condition that the ranges of available
     $x_e$ and $x_i$ are large compared to the variation range of
     one of them at a fixed value of the other variable. We have used
     $m_e / m_i = 0.2$, $\eta = 0.5$, and $\gamma_I t = 4$ for illustration.
     The dashed lines drawn in the figure tell approximately the region of
     localization of the wave function.}
\label{fig2}
\end{figure}

Now we focus on measurements appropriate for seeing entanglement.
We need to distinguish coincidence and non-coincidence
(single-particle) measurements, which have their theoretical
counterparts in conditional and non-conditional probability
distributions.  For an example of a single-particle measurement,
the electron probability distribution is measured regardless of
the ion position (or vice versa).  In contrast, a coincidence
measurement assumes that a distribution of electron positions is
registered while the ion detection position is kept at a given
(constant) location (or vice versa). The difference between the
results of coincidence and single-particle schemes of measurements
is illustrated by Fig.~\ref{fig2}.  In this picture, in one
dimension, we shade the region in which the joint probability
density $|\Psi(\vec{r}_e,\vec{r}_i,t)|^2$   is significant. In the
left plot the sharp leading edge of the theta function in
Eq.~(\ref{Psi-residue-int}) is apparent, with its long exponential
tail, and one also sees the more abrupt Gaussian cut-off on the
sides. A purely schematic view of the same thing is shown in the
right plot, where artificially sharp dashed-line borders are
introduced, and supposed to be determined by the localization
zones of the relative-motion and center-of-mass wave functions.

Consider first an examination of the electron wave packet by the
coincidence-scheme method, for a given ion coordinate $x_i =
\text{const}$. The normalized measure of its width, $\Delta
x_e/\Delta r_\text{rel}(t)$, will be given by the distance between
the points marked $a$ and $b$.  In contrast, the single-particle
width takes into account the contributions from all possible
different $x_i$'s. Thus a suitable measure of the single-particle
width of the electron wave packet is given by the distance  $cd$.
It is obvious that the electron packet is relatively highly
localized when $cd \gg ab$. Correspondingly, a horizontal line
through the shaded region would provide a normalized measure of
$\Delta x_i$, etc. From this sketch we formulate  two conditions
simultaneously necessary for entanglement to be large: a high
aspect ratio of the shaded area and a nearly diagonal angle
between the dashed lines restricting the wave packet localization
zones and the coordinate axes $x_e$ and $x_i$. The high aspect
ratio condition means that one of the two wave packets (${\rm
``cm"}$ or ${\rm ``rel"}$) is much wider than the other one.

Now we note that this relatively great localization condition is
the same as a high entanglement condition. This becomes obvious by
considering the two-particle wave packet
$|\Psi(\vec{r}_e,\vec{r}_i,t)|^2$   as an information container
\cite{Grobe-etal}. Then the bibliography gets the added element:

Entanglement means that knowledge of one of the particles imparts
information about the other, whereas non-entangled particles
provide no information about each other. The greatest
cross-specification of joint information by an entangled packet
occurs when knowledge of one particle automatically corresponds to
precise information about the other, i.e., knowledge of position
$x_i$ strongly localizes the region where $x_e$ can be found.
Reflection shows that a thin diagonal packet in $x_i$-$x_e$ space
achieves this, and normalized relative information gain is well
expressed by the ratio of single to coincidence width. Note that
left or right inclination of the diagonal is immaterial. In
contrast, in wave packet language the non-entangled condition
(information about $x_i$ gives no information about $x_e$) is
equivalent to a factored wave packet:   $\Psi(x_i, x_e) \to
\phi_1(x_i)~\phi_2(x_e)$. One sees that a  sketch corresponding to
Fig.~\ref{fig2}, but for independent particles, has dashed lines
that are horizontal and vertical, in which case all $x_i$'s
predict exactly the same $\Delta x_e$, and vice versa.

Mathematically, single-particle probability densities are
given by Eq.~(\ref{product-2}) integrated either
over $\vec{r}_i$ or $\vec{r}_e$:
\begin{equation}
\label{single-el}
     P_e(\vec{r}_e, t) = \int d\vec{r}_i \ |\Psi(\vec{r}_e,
\vec{r}_i, t)|^2,
\end{equation}
or
\begin{equation}
\label{single-i}
     P_i(\vec{r}_i, t) = \int d\vec{r}_e \ |\Psi(\vec{r}_e,
\vec{r}_i, t)|^2.
\end{equation}
Such distributions reveal no entanglement effects because all the information
about the position of one of the particles is lost completely when the
two-particle probability density is integrated over $\vec{r}_i$ or $\vec{r}_e$.
However, $P_e$ and $P_i$ do serve a normalization role, as we explain later.

Let $\Delta r_e^\text{(s)}$ and $\Delta r_i^\text{(s)}$ be the widths
of the single-particle electron and ion wave packets, where, for example,
$|\Delta r_e^\text{(s)}|^2
= \langle|\vec{r}_e|^2\rangle-|\langle\vec{r}_e\rangle|^2$, with
\begin{eqnarray}
     \langle \vec{r}_e \rangle
     &=& \int d\vec{r}_e \ \ \vec{r}_e \ P_e(\vec{r}_e, t)
     \nonumber \\
     &=& \iint d\vec{r}_e \, d\vec{r}_i \ \ \vec{r}_e \
|\Psi(\vec{r}_e, \vec{r}_i, t)|^2
     \nonumber \\
     &=& \langle\vec{r}_\text{cm}\rangle
         + \frac{m_i}{M}\langle\vec{r}_\text{rel}\rangle
\label{eq:mean-re}
\end{eqnarray}
and
\begin{eqnarray}
     \langle \left|\vec{r}_e\right|^2 \rangle
     &=& \int d\vec{r}_e \ \ r_e^2 \ P_e(\vec{r}_e, t)
     \nonumber \\
     &=& \iint d\vec{r}_e \, d\vec{r}_i \ \ r_e^2 \
|\Psi(\vec{r}_e, \vec{r}_i, t)|^2
     \nonumber \\
     &=& \langle\left|\vec{r}_\text{cm}
         + \frac{m_i}{M}\vec{r}_\text{rel}\right|^2\rangle
     \nonumber \\
     &=& \langle\left|\vec{r}_\text{cm}\right|^2\rangle
         + 2\frac{m_i}{M}
\langle\vec{r}_\text{cm}\rangle\langle\vec{r}_\text{rel}\rangle
         +
\frac{m_i^2}{M^2}\langle\left|\vec{r}_\text{rel}\right|^2\rangle .
     \hspace{8mm}
\label{eq:mean2-re}
\end{eqnarray}
Note that we have used the relation
$\vec{r}_e = \vec{r}_\text{cm} + \frac{m_i}{M} \vec{r}_\text{rel}$
and changed the integration variables to the center-of-mass and relative
coordinates.  Then Eqs.~(\ref{eq:mean-re}) and~(\ref{eq:mean2-re})
yield the single-particle measures:
\begin{equation}
\label{Delta-r-e-s}
     \delta r_e^\text{(s)}
     \equiv \frac{\Delta r_e^\text{(s)}}{\Delta r_\text{rel}(t)}
     = \sqrt{\eta^2(t)+\left(\frac{m_i}{M}\right)^2} ,
\end{equation}
and, similarly,
\begin{equation}
\label{Delta-r-i-s}
     \delta r_i^\text{(s)}
     \equiv \frac{\Delta r_i^\text{(s)}}{\Delta r_\text{rel}(t)}
     = \sqrt{\eta^2(t)+\left(\frac{m_e}{M}\right)^2} .
\end{equation}
Note that the relative-motion wave packet width $\Delta r_\text{rel}(t)$
plays the role of a natural normalization factor for both single-particle
and coincidence-scheme (see below) electron and ion wave packet widths.
Divided by $\Delta r_\text{rel}(t)$ these widths become dimensionless,
and they are denoted $\delta r_{e,i}^\text{(s)}$ and
$\delta r_{e,i}^\text{(c)}$.

In the coincidence scheme of measurements
the overall width of the distribution (\ref{product-2})
with respect to $\vec{r}_e$ at a fixed $\vec{r}_i$ is given by
the smaller of $(M/m_e) \Delta r_\text{cm}(t)$
and $\Delta r_\text{rel}(t)$, which is well-represented by a simple
formula for the coincidence measures:
\begin{equation}
\label{Delta-r-el-coinc}
     \delta r_e^\text{(c)}
     \equiv \frac{\Delta r_e^\text{(c)}}{\Delta r_\text{rel}(t)}
     \approx \frac{\eta(t)}
     {\sqrt{\eta^2(t) + \left(\displaystyle\frac{m_e}{M} \right)^2}}.
\end{equation}
The expression in Eq.~(\ref{Delta-r-el-coinc})
is a better approximation the closer we are to one of the extreme cases
$(M/m_e)\Delta r_\text{cm}(t) \gg \Delta r_\text{rel}(t)$ or
$\ll \Delta r_\text{rel}(t)$.
As shown later in this section, these limits correspond to the high
entanglement
regimes of main interest, and we do not need to bother too much about
the details of the intermediate region.  Similarly, at a given
$\vec{r}_e$, the widths of $|\Psi_\text{cm}|^2$ and
$|\Psi_\text{rel}|^2$ with respect to $\vec{r}_i$ are correspondingly
$(M/m_i)\Delta r_\text{cm}(t)$ and $\Delta r_\text{rel}(t)$.
Therefore the overall width of $|\Psi|^2$ at a given $\vec{r}_e$ is
the smaller of $(M/m_i)\Delta r_\text{cm}(t)$ and $\Delta r_\text{rel}(t)$,
corresponding to the formula
\begin{equation}
\label{Delta-r-i-coinc}
     \delta r_i^\text{(c)}
     \equiv \frac{\Delta r_i^\text{(c)}}{\Delta r_\text{rel}(t)}
     \approx \frac{\eta(t)}
     {\sqrt{\eta^2(t) + \left(\displaystyle\frac{m_i}{M} \right)^2}} .
\end{equation}

\begin{figure}[!htb]
     \centering\epsfig{file=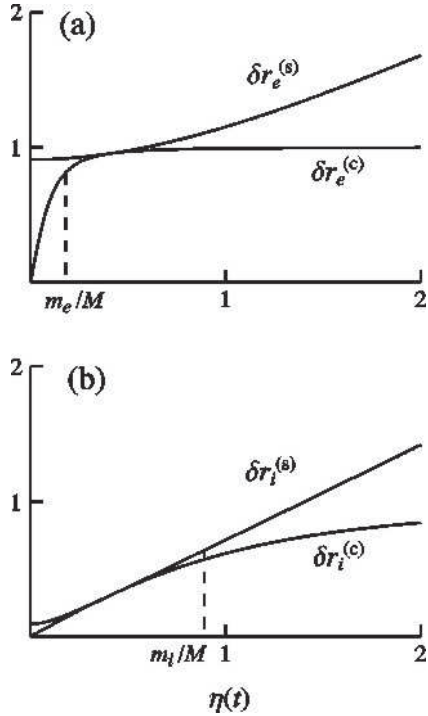, width=5.5cm}
     \caption{Electron (a) and ion (b) wave-packet widths in
     the schemes of single-particle and coincidence measurements
     with $m_e/m_i=0.1$.}
\label{fig3}
\end{figure}

It is clear now how $\eta(t)$ of Eq.~(\ref{eq:eta}) serves as a
``control parameter'' for both coincidence widths. Plots of
$\delta r_i^\text{(s)}$ and $\delta r_i^\text{(c)}$ as a function
of $\eta$, are shown in Fig.~\ref{fig3}(a) whereas graphs of
$\delta r_i^\text{(s)}$ and $\delta r_i^\text{(c)}$ are shown in
Fig.~\ref{fig3}(b). Note that we use in these graphs an artificial
value of the electron to ion mass ratio $m_e/m_i=0.1$ so as to
show more clearly the difference between the two curves for $\eta
< 1$. However, all the qualitative conclusions from these pictures
remain the same for a more realistic value of this ratio
$m_e/m_i\sim 10^{-4}$. One of these conclusions is that we always
have $\delta r_{e, i}^\text{(s)} > \delta r_{e, i}^\text{(c)}$.

The ratios of single-to-coincidence electron and ion wave packet widths,
$\delta r_{e,i}^\text{(s)}$ to $\delta r_{e,i}^\text{(c)}$,
can be considered as a measure of entanglement, as remarked at the
beginning of this section. As we will show elsewhere
\cite{laterpaper}, they are essentially identical to the
corresponding Schmidt number discussed in earlier discussions of
photon-atom entanglement \cite{JHE2,JHE}. These ratios can be
referred to as the electron and ion entanglement parameters in the
form:
\begin{equation}
\label{ent-param}
     R_e
     \equiv \frac{\delta r_e^\text{(s)}}{\delta r_e^\text{(c)}}
     \quad\text{and}\quad
     R_i
     \equiv \frac{\delta r_i^\text{(s)}}{\delta r_i^\text{(c)}}.
\end{equation}
Entanglement is large if $R_e\gg 1$ and/or $R_i\gg 1$. If
$R_e\approx 1$ and $R_i\approx 1$, there is little or no entanglement
at all.

By using Eqs.~(\ref{Delta-r-e-s}) -- (\ref{Delta-r-i-coinc}), we can find
a useful approximate form of $R_e$ and $R_i$ :
\begin{equation}
     \label{R(eta)}
       R_e = R_i
       \approx \sqrt{\eta + \frac{1}{\eta}\left(\frac{m_i}{M}\right)^2}
       \sqrt{\eta + \frac{1}{\eta}\left(\frac{m_e}{M}\right)^2} ,
\end{equation}
which is plotted in  Fig.~\ref{fig4}. Note that here the width
ratios for the electron and the ion are the same, i.e., $R_e =
R_i$.  This is actually true only when the widths of
$\Psi_\text{cm}$ and $\Psi_\text{rel}$ are very different from
each other, or equivalently $\eta \gg 1$ or $\eta \ll 1$.  Even
though $R_e$ and $R_i$ may not be exactly the same in the zone
$m_e/M < \eta < 1$, they both have values around unity, which
corresponds to the relatively less interesting small entanglement
regime. Thus we designate the two of them together by $R$ without
a subscript. The asymptotic behaviors of $R$ in three different
regions of $\eta(t)$ are particularly noteworthy:
\begin{eqnarray}
     \text{Region 1}, \ \
     \eta \ll \frac{m_e}{M} \ll \displaystyle\frac{m_i}{M}:
     & & \nonumber \\
     & & \hspace{-25mm}\label{reg1}
     R \sim \frac{m_e m_i}{M^2}
     \frac{\Delta r_\text{rel}(t)}{\Delta r_\text{cm}(t)}
     \sim \left(\frac{\mu}{M}\right) \frac{1}{\eta(t)} ;
     \qquad
     \\[3mm]
     \text{Region 2}, \ \
     \frac{m_e}{M} \ll \eta \ll \frac{m_i}{M}:
     & &
     R \sim 1\label{reg2} ;
     \\[3mm]
     \text{Region 3}, \ \
     \frac{m_e}{M} \ll \frac{m_i}{M} \ll \eta :
     & &
     R \sim \eta(t)\label{reg3} .
\end{eqnarray}

Note that the minimal value of the entanglement parameter
(\ref{ent-param}) is equal to one, $R_{\min} =1$, and it is
achieved at $\eta(t)=\sqrt{\mu/M}$.

\section{Time Evolution of Packet Widths and Entanglement Parameter}
\label{section:evolution}

\begin{figure}[!htb]
     \centering\epsfig{file=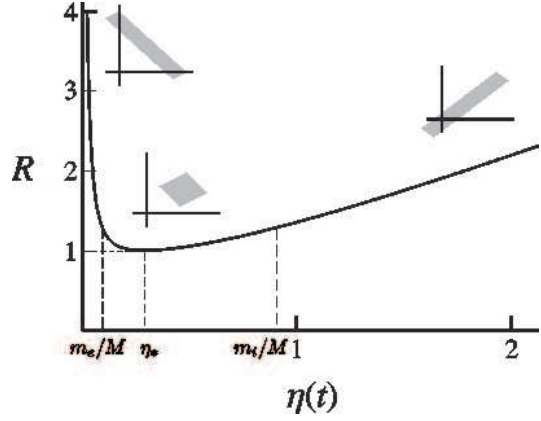, width=7cm}
     \caption{Plot of the entanglement parameter $R$ as a function of
     $\eta(t)$ with $m_e / m_i = 0.1$; $\eta_*$ is the stability point
     (\ref{eta*}) at which $\eta(t)\equiv const=\sqrt{\mu/M}$.
     The insets give the corresponding plots of the one-dimensional
     analog $|\Psi(x_e,x_i,t)|^2$ from Fig.~\ref{fig2}.
     The axes of the three insets have been rescaled so as
     to show the details more clearly.  The large entanglement regions
     are clearly seen to correspond to a large aspect ratio of the
     shadowed areas.}
\label{fig4}
\end{figure}

In Figs.~\ref{fig3} and~\ref{fig4} both the electron/ion wave
packet widths and the entanglement parameters are shown, for fixed $t$,
in their dependence on the control parameter defined earlier:
\begin{equation} \label{eq:etaDef}
    \eta(t)  = \frac{\Delta r_\text{cm}(t)}{\Delta r_\text{rel}(t)} .
\end{equation}
However, we can use the same pictures to show the time
evolution of the widths $\Delta r_{e}(t)$ and $\Delta r_{i}(t)$
and the entanglement parameter $R(t)$ defined in (\ref{R(eta)}). To do this,
we have to learn how $\eta(t)$ changes with time.

\begin{figure}[!htb]
     \centering\epsfig{file=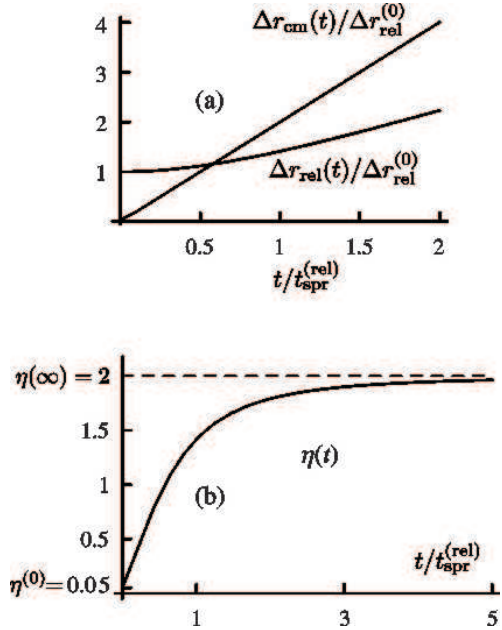, width=6.5cm}
     \caption{Part (a) shows the time-dependent widths of the center-of-mass
     and relative-motion wave packets (in units of $\Delta
     r_\text{rel}^{(0)}$), and part (b) shows the control parameter
     $\eta(t)$.  We have taken $\eta_0=0.05$ and $m_e/M=0.1$.}
\label{fig5}
\end{figure}

\begin{figure}[!htb]
     \centering\epsfig{file=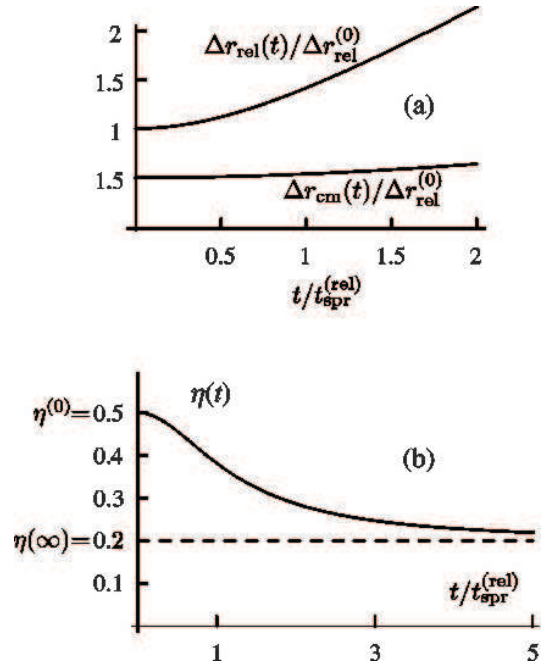, width=7cm}
     \caption{The same as in Fig.~\ref{fig5} but with
     $\eta_0=0.5$.}
\label{fig6}
\end{figure}

The two typical cases of significantly different behavior
are illustrated in Figs.~\ref{fig5} and~\ref{fig6}. Parts (a) and (b) of these
Figures show the time dependence of
the widths $\Delta r_{\text{cm}}(t)$ and $\Delta
r_{\text{rel}}(t)$ themselves and of their ratio, which equals $\eta(t)$.
A key feature of $\eta(t)$ is its strong dependence on
the initial sizes of the center-of-mass and relative-motion wave
packets, $\Delta r_{\rm cm}^{(0)}$ and $\Delta r_{\rm rel}^{(0)}$,
or in other words, on the initial value of the control parameter
$\eta(0) \equiv \eta_0$. Depending on its initial value,
$\eta$ is either rising
as shown in Fig.~\ref{fig5}(a) or falling as in Fig.~\ref{fig6}(b).
The border between these two regimes is given by $\eta_0=\eta_*$, where
\begin{equation} \label{eta*}
  \eta_* \equiv \sqrt{\displaystyle\frac{\mu}{M}}.
\end{equation}

If $\eta_0<\eta_*$, the center-of-mass wave packet spreads faster
and eventually becomes wider than the relative-motion wave packet,
though initially $\Delta r_{\rm cm}^{(0)}\ll\Delta r_{\rm
rel}^{(0)}$. In this case the control parameter $\eta(t)$ is a
monotonically growing function of $t$ (see Fig.~\ref{fig5}(b)).
On the contrary, if $\eta_0>\eta_*$, the center-of-mass wave packet
spreads slower than the relative-motion wave packet. Though
initially the center-of-mass packet can be either narrower or
wider than the relative-motion packet, at
very large $t$ the relative-motion packet becomes wider than
the center-of mass packet. This gives rise to a falling function
$\eta(t)$ shown in Fig.~\ref{fig6}(b). In both cases ($\eta_0<\eta_*$
and $\eta_0>\eta_*$) the ranges of variation of the parameter
$\eta(t)$ are finite. At very long times  $\eta(t)$
has the asymptotic value
\begin{equation} \label{eta-asympt}
  \eta \to \eta_\infty \equiv \frac{\mu}{M}\frac{1}{\eta_0},
\end{equation}
which follows directly from the definition of $\eta$ (\ref{eq:etaDef})
and Eqs.~(\ref{spr-width}) and~(\ref{Detlta-rel-tot}) for the
widths $\Delta r_{\text{cm}}(t)$
and $\Delta r_{\text{rel}}(t)$. In the case $\eta_0=\eta_*$ the
parameter $\eta(t)$ does not depend on time at all:
$\eta(t) = const ~(=\eta_*=\eta_0=\eta_\infty)$.

By finding the evolution regimes for the control parameter
$\eta(t)$, we can also make definite and interesting
conclusions on the evolution of the entanglement parameter $R(t)$
(\ref{R(eta)}). Directly from Eq. (\ref{R(eta)}) one can easily see
that for an arbitrary value of $\eta$, the
entanglement parameter obeys the relation
\begin{equation} \label{R(1/eta)}
    R\left(\frac{1}{\eta}\;{\frac{\mu}{M}}\right) \equiv R(\eta).
\end{equation}
The initial and final values of $\eta$ are connected with each other
exactly by the
same substitution as used in Eq. (\ref{R(1/eta)}), we see that
the initial and final values of the entanglement parameter must be equal:
\begin{equation}
  \label{R-infty}
  R_0\equiv R_\infty =R(t\rightarrow\infty).
\end{equation}
For the entanglement parameter given by Eq. (\ref{R(eta)}) this
equality is valid identically for all values of $\eta_0$.
If $\eta_0$ is located in one of the high-entanglement regions of
Fig.~\ref{fig4}, $\eta_0\ll\mu/M$ as in (\ref{reg1}), or
$\eta_0\gg 1$ as in (\ref{reg3}), the final value of $\eta$ is in the
opposite of
these two high-entanglement regions, $\eta_\infty\gg 1$ or
$\eta_\infty\ll\mu/M $.

Thus we see that the time-dependent entanglement parameter
$R(t)$ starts from a large value $R_0$, falls to $R\sim 1$ and
then grows again to the same value from which it started.
Physically such an evolution means that initially one of the wave
packets is much wider than the other one, and
for this reason the electron-ion entanglement is
large. Then, as the narrower wave packet spreads faster, they
become approximately of the same width, and this corresponds to a
small entanglement. Finally, when the initially narrower but
faster spreading wave packet outstrips the initially wider but
slower spreading one, the relation between their widths reverses,
and this returns us to the case of a large entanglement.

The difference between the cases $\eta_0\ll\mu/M$ and $\eta_0\gg
1$ concerns only the direction of evolution, correspondingly, to
the right or to the left in the $\eta$-axis in Fig.~\ref{fig4}. If
the initial value of the parameter $\eta$ is located in the
small-entanglement region (\ref{reg3}), $\mu/M<\eta<1$, all the
conclusions about the direction of evolution and about the
relation between the initial and final values of the entanglement
parameter remain valid. However, in this case, at all times $t$
the entanglement parameter remains on the order of one. If
$\eta_0=\eta_*$, the entanglement parameter does not change at
all, and $R(t)\equiv 1$.

\section{Experimental Considerations}

\begin{figure}[!htb]
     \centering\epsfig{file=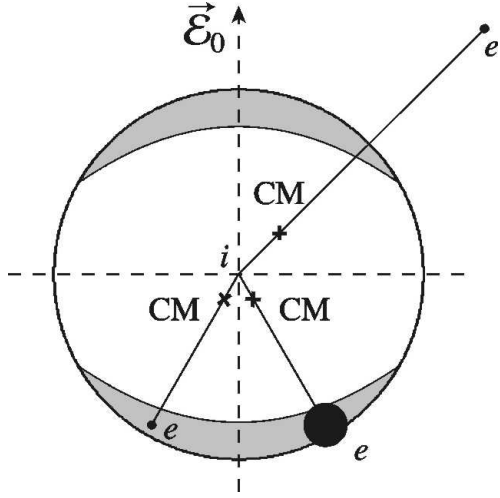, width=6.5cm}
     \caption{The new-moon shaded areas indicate the regions where
the function $|\Psi_\text{rel}(\vec{r}_e)|^2$ is relatively
large,relatively large, i.e., at least as large as one-third the
maximum value. Black dots indicate regions where
$|\Psi_\text{cm}(\vec{r}_e)|^2\neq 0$, at a given $\vec{r}_i$.
Three shown experimental situations correspond to different
locations of the electron detector, relative to the ion position,
which defines the origin. A smaller size of such a black dot
inside the shaded area corresponds to a higher level of
entanglement. In the case when the black dot is located far
outside of the shaded area, there is no overlapping between the
center-of-mass and relative-motion wave functions and the total
two-particle wave function equals zero, $\Psi=0$. } \label{fig7}
\end{figure}

The discussion as given so far doesn't treat some elements that
will come into play in experimental tests. In order to bring them
into focus briefly, we show in Fig.~\ref{fig7} what can be called
experimentally realistic zones. We have plotted the region where
the relative probability distribution is non-zero. It has a
three-dimensional aspect that we do not need to show because it is
axially symmetric about the polarization axis of the ionizing
light beam, taken as the vertical axis here. It is not spherically
symmetric because of the dipole character of photoionization
(i.e., the factor $\vec{d}_{E_*0} \cdot \vec{\mathcal{E}}_0$ in
Eq.~(\ref{Psi-residue-int})). The new-moon shaded areas indicate
the regions where the relative-motion wave function
$|\Psi_\text{rel}(\vec{r}_e)|^2$ is relatively large.

Since the time evolution of the relative wave function is strictly
limited by the step function $\theta(vt - r_\text{rel})$, we will
here consider the ion position to define an origin of polar
coordinates ($r_i \equiv 0$), in which case a circle of radius
$r_e = vt$ limits the range of the electron coordinate at time
$t$. The relative coordinate probability distribution
$|\Psi_\text{rel}(\vec{r}_e)|^2$ is of course not uniform inside
this circle, so we have drawn the boundary on which
$|\Psi_\text{rel}|^2$ equals $\frac{1}{3}$ of its maximum value.
This creates two sectors with ``new-moon" shape where there is the
highest probability to find the electron, given that the ion is at
the origin of the circle, and taking only $|\Psi_\text{rel}|^2$
into account. However, the probable position of the electron is
also influenced by $|\Psi_\text{cm}(\vec{r}_e)|^2$ at
$\vec{r}_i\equiv 0$. In the figure the black dot regions show the
range of electron positions given by
$|\Psi_\text{cm}(\vec{r}_e)|^2$ at different positions of the
center of mass along the $r_e$-$r_i$ line.

The figure has many variations, and the sizes of the new-moon and
$\Delta r_e$ zones change in time, as our formulas indicate. The
overall shapes will remain the same, and a generic high-entanglement
experiment will be one that ensures overlap between a small black-dot
region and a high-probability portion of one of the new-moon regions
(near the circle boundary). Higher count rate, although reduced
entanglement, will be associated with increased size of the black dot
region.

\section{Photodissociation}

\begin{figure}[!htb]
     \centering\epsfig{file=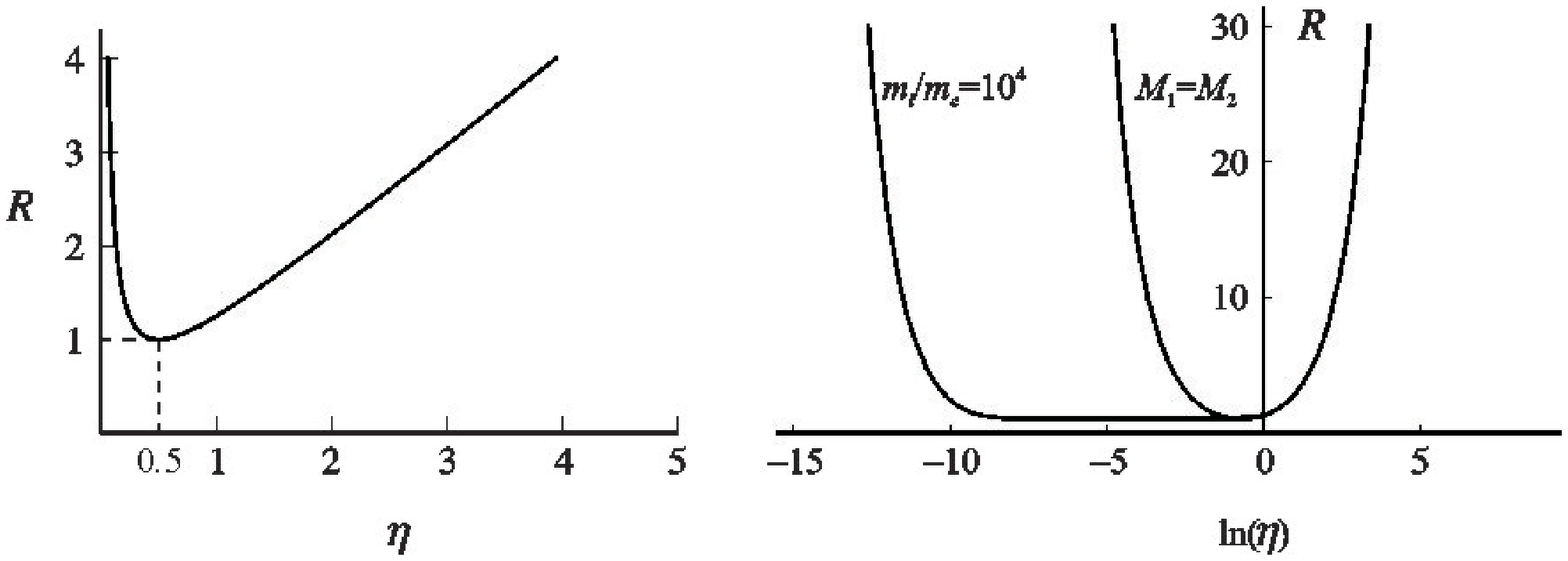, width=13cm}
     \caption{Entanglement parameter for two dissociating molecular fragments
     with $M_1=M_2$ (left) and the same dissociation curve plotted vs. ${\rm
     ln}(\eta)$ on the right, where the corresponding photoionization curve is
     included for comparison, with its very different mass ratio, $m_i =
     10^4 m_e$. } \label{fig8}
\end{figure}

It is easy to see that very similar results will arise in a
treatment of photodissociation of molecules. Here we remark
briefly on some of the differences. Let us assume that we consider
a diatomic molecule undergoing dissociation. There will be a
relevant dissociation rate $\gamma_D$, which can be substituted
for the $\gamma_I$ governing ionization, and just as for the atom
there will be an initial localization of the molecular center of
mass. Then the main differences to the ionization example arise
because the mass ratio of the fragments is much closer to 1.
Compared to the case of photoionization, where $m_e \ll m_i$, the
masses $M_1$ and $M_2$ of the photodissociation fragments obey
$M_1\sim M_2$. Given this, the relative-motion velocity after
dissociation $v \sim \sqrt{\hbar\omega/\mu}$ is significantly
smaller than in the case of ionization (in atomic units
$v_\text{mol} \sim \sqrt{\omega/M}$). The main difference between
photoionization and photodissociation results concerns the region
$\frac{\mu}{M} < \eta(t) < 1$ of intermediate values of $\Delta
r_\text{cm}(t)$ in Fig.~\ref{fig4} where $R\sim 1$. For
$M_1=M_2=\frac{1}{2}M$ this region degenerates into a single point
$\eta(t)=\frac{1}{2}=\eta_*$ (\ref{eta*}). The entanglement coefficient
$R$ is large both at $\eta(t)<\eta_*$ and $\eta(t)>\eta_*$. To
show more clearly the difference between photoionization and
photodissociation we plot in the right picture of Fig.~\ref{fig8}
both molecular and atomic entanglement coefficients in their
dependence on ${\rm ln}(\eta)$, with the electron to ion mass
ration taking a realistic value $m_e/m_i=10^{-4}$. This picture
shows that if in the case of photoionization there is a rather
large region of intermediate values of $\eta$ where entanglement
is small, $R\approx 1$, in the case of photodissociation of a
molecule the entanglement parameter is large practically at any
$\eta$ except one point $\eta=\eta_*$.

In dependence on time $t$ the control parameter $\eta(t)$ changes
in a way similar to that described above for photoionization:
$\eta(t)$ grows if initially it is small ($\eta_0<\eta_*$) and falls
if large ($\eta_0>\eta_*$). The final value of the
control parameter $\eta_\infty$ is related to $\eta_0$ by Eq.
(\ref{eta-asympt}), which takes the form $\eta_\infty=1/(4\eta_0)$.
As shown previously, the initial and final values of the time-dependent
entanglement parameter $R(t)$ are equal to each other, $R_0=R_\infty$.
At $\eta_0=\eta_*$ both the control parameter $\eta(t)$ and the
entanglement parameter $R(t)$ do not change with a varying
time $t$, $\eta(t)\equiv\eta_0$ and $R(t)\equiv 1$. This is the
only case when there is no entanglement at any time $t$. In all
other cases ($\eta_0\neq\eta_*$) the entanglement parameter is
large initially, reaches $R=1$ one at such $t$ that gives
$\eta(t)=\eta_*$, and then grows again until it reaches its
initial value $R_0$.

\section{Conclusion}

We have evaluated the space-time behavior of the joint quantum state
of an ion and electron following photoionization. Neglect of the
incident photon momentum and of the final state Coulomb interaction means
that the evolution of the
state, and thus of the entanglement between the two particles, is
constrained only by free-particle two-body momentum and energy
conservation. This evolution provides an exactly calculable illustration
of the situation involving massive particles sketched in the famous
paper of Einstein, Podolsky, and
Rosen~\cite{EPR}. We have obtained expressions for the
entanglement-induced wave
packet narrowing that occurs, and have indicated how  entanglement can be
identified and determined quantitatively. To do this we introduced
$R$, the ratio between the
entanglement-free wave packet width and the coincidence wave packet
width. This is essentially the degree of entanglement. We gave
expressions for $R$ in
terms of ionization rate and packet spreading velocity, which are of
course themselves determined by underlying parameters such as atomic
bound-free dipole moments,
relative electron and ion masses, ionizing field strength, etc.  It was shown
that $R$ depends in a simple way on the basic control parameter $\eta =
\Delta r_\text{cm}(t)/\Delta r_\text{rel}(t)$, and can be much larger
than unity in  two limits, when $\eta \gg 1$ and also $\eta \ll 1$.  The
same formalism can be applied equally well to photodissociation of a
diatomic molecule. For realistic physical values of the relevant
parameters, in a typical example of atomic photoionization,
$R$ is not very large because of
the extreme discrepancy between $m_i$ and $m_e$, but for
photodissociation of a diatomic molecule, where the fragment masses can
be approximately equal, $R$ can be substantially increased.

\section{Acknowledgements} The research reported here has been
supported by the DoD Multidisciplinary University Research Initiative
(MURI) program administered by the Army Research Office under Grant
DAAD19-99-1-0215, by NSF grant PHY-0072359, Hong Kong Research Grants Council
(grant no. CUHK4016/03P), and the RFBR grant 02-02-16400.


\begin{thebibliography}{99}
\bibitem{laterpaper} Universal rules for quantification and
measurement of bipartite entanglement, explicitly joining the Schmidt
context to localization measurements, will be treated in a wider
context elsewhere.
\end{thebibliography}
\end{document}